\documentclass{article}

\usepackage{amsmath,amssymb,amsfonts}
\usepackage{algorithmic}
\usepackage{graphicx}
\usepackage{textcomp}
\usepackage{xcolor}
\usepackage{balance}
\usepackage{enumitem}
\usepackage{tcolorbox}
\usepackage{multirow}

\AtBeginDocument{%
  }

\usepackage{arxiv}

\usepackage[utf8]{inputenc} 
\usepackage[T1]{fontenc}    
\usepackage{hyperref}       
\usepackage{url}            
\usepackage{booktabs}       
\usepackage{amsfonts}       
\usepackage{nicefrac}       
\usepackage{microtype}      
\usepackage{lipsum}
\usepackage{graphicx}
\graphicspath{ {./images/} }

\title{Insights into Security-Related AI-Generated Pull Requests}

\author{
 Md Fazle Rabbi \\
  Idaho State University, USA \\
  \texttt{mdfazlerabbi@isu.edu} \\
   \And
 Asif K. Turzo \\
  University of Massachusetts Dartmouth, USA \\
  \texttt{aturzo@umassd.edu} \\ 
  \And
 Arifa I. Champa \\
  Idaho State University, USA \\
  \texttt{arifaislamchampa@isu.edu} \\
  \And
 Minhaz F. Zibran \\
  Idaho State University, USA \\
  \texttt{zibran@isu.edu} \\
}

\begin{document}
\maketitle
\begin{abstract}

Recent years have experienced growing contributions of AI coding agents that assist human developers in various software engineering tasks. However, this growing AI-assisted autonomy raises questions about security and trust. 
%
In this paper, we analyze more than 33,000 AI-generated pull requests (PRs) and identify 675 security-related submissions made by agentic AIs. Then we examine the security-related PRs with a focus on recurring security weaknesses, review outcomes and latency, commit message quality, and rejection reasons. The results show that security-related AI PRs introduce a small set of recurring weaknesses such as regex inefficiencies, injection flaws, and path traversal. Many flawed contributions are still merged, while rejections often arise from social or process factors such as inactivity or missing test coverage. The commit message quality of AI PRs has a limited effect on acceptance or latency, in contrast to human PRs reported in previous studies. We also extend existing rejection taxonomies by adding categories that are unique to AI-generated security contributions. These findings offer new insights into the strengths and shortcomings of autonomous coding systems in secure software development.  
\end{abstract}


\section{Introduction}

Software security has become one of the most pressing challenges in modern software engineering~\cite{bohme2025software, assal2025software}. Open-source (OS) projects depend on a continuous flow of contributions to fix vulnerabilities, update dependencies, and patch flaws~\cite{lin2024vulnerabilities, yang2025code}. Traditionally, human developers submit these contributions through pull requests (PRs)~\cite{gousios2014exploratory}. Reviewers then examine the proposed changes and decide whether to merge them. This process has long been central to collaborative software development~\cite{6606617}.

The emergence of large language models (LLMs) is beginning to reshape this process~\cite{chouchen2024software,xiao2024generative}. LLMs are no longer restricted to providing code completion or acting as assistants within an editor. They now generate full commits and submit PRs automatically~\cite{watanabe2025use}. This practice, often referred to as agentic coding, is increasingly common in open-source communities~\cite{sapkota2025vibe}. 

The rise of AI-generated PRs brings both opportunities and risks. On the positive side, automated contributions can reduce the delay between vulnerability disclosure and patch submission~\cite{dissanayake2022empirical}. They can manage repetitive and large-scale tasks, such as dependency upgrades, that are difficult for human contributors to sustain. On the negative side, AI-generated code may lack context or awareness of system-specific requirements. Subtle errors in a patch can weaken security instead of improving it. Reviewers must decide whether to accept or reject such changes~\cite{Ogenrwot2024PatchTrack}, often without clear visibility into the reliability of the underlying AI agent.

Recent work has started to examine how open-source communities react to AI-generated PRs. Some studies report cautious trust and selective acceptance of AI-generated contributions, such as those created by Claude Code~\cite{watanabe2025use} or ChatGPT~\cite{Ogenrwot2024PatchTrack}. However, most of these studies focus on general-purpose PRs. Security-related PRs stand apart from general code contributions because they usually involve small but highly sensitive edits. These changes often touch critical areas such as cryptographic routines, configuration files, or dependency versions linked to known CVEs~\cite{rebatchi2024dependabot,alfadel2021use}. Even a minor mistake in such edits can expose an application to new security risks instead of fixing existing ones. As a result, reviewers tend to examine these PRs more carefully and apply stricter standards before approval~\cite{stiefel2025securityprs}. Because of their potential impact, security-related AI PRs need to be studied separately rather than treated as part of general AI-generated contributions.

We address this gap by conducting the first large-scale analysis of how agentic AI systems contribute to software security processes. We analyze more than 33,000 AI-generated PRs from the AIDev dataset~\cite{li2025rise} and construct a subset focused on security-related contributions. Using this dataset, we examine the types of security issues involved, factors influencing review latency and acceptance, the quality of commit messages, and the reasons for rejection. Our investigation addresses the following research questions (RQs):


\noindent \textbf{RQ1: What type of security weaknesses are introduced by security-related AI PRs?}

\noindent \textit{Rationale:} Security-related PRs submitted by AI agents can sometimes add new vulnerabilities instead of fixing existing ones. It is important to understand what kinds of weaknesses appear most often and which AI agents tend to cause them, so that both project maintainers and AI developers can improve the security and reliability of future contributions.

\noindent \textbf{RQ2: Which factors are associated with the review latency and acceptance outcomes of security-related AI PRs?}

\noindent \textit{Rationale:} Reviews of security-related AI PRs may differ in speed and outcome, yet little is known about the factors behind these variations. Identifying which project-level features influence latency and acceptance can help dealing with AI contributions more effectively and help developing more reliable coding agents.

\noindent \textbf{RQ3: What is the quality of commit messages in security-related AI PRs, and how does it affect review outcomes?}

\noindent \textit{Rationale:} Commit messages help reviewers judge the intent and clarity of a change, yet AI agents may differ in how well they generate them. Understanding which agents produce higher or lower quality messages and how this relates to PR acceptance can show whether message clarity influences trust in AI-generated security contributions.

\noindent \textbf{RQ4: Why are security-related AI PRs rejected?}

\noindent \textit{Rationale:} Many security-related AI PRs are rejected, but the reasons behind these decisions remain unclear. Identifying why maintainers reject such PRs and whether the causes differ across AI agents can reveal the shortcomings in current review practices and highlight areas where AI systems and workflows need improvement.

We address these questions through a combination of quantitative and qualitative analyses. Our results show that certain AI agents repeatedly introduce similar types of vulnerabilities, review outcomes are shaped by project-level characteristics, and rejection patterns differ across agents. This study makes \textbf{four main contributions}:

\vspace{-0.1cm}
\begin{itemize}

    \item We detect vulnerabilities introduced by security-related AI PRs and report the types and frequency of security issues identified across diverse open-source projects.

    \item We analyze 24 factors derived from prior PR studies and adapt them to the AI security context. Using regression models, we identify factors that significantly influence review latency and acceptance. We also evaluate commit message quality and its relationship to review outcomes.

    \item We conduct a detailed analysis of rejection reasons for security-related AI-generated  PRs. Building on established rejection taxonomies~\cite{pantiuchina2021developers, watanabe2025use}, we identify common causes of rejection and compare patterns across AI agents.

    \item We compile the dataset of \emph{security-related} AI-generated PRs which links PR metadata, repository context, and structured security labels. To support replication, we make our code and dataset publicly available at~\cite{dataset}.

\end{itemize}
\vspace{-0.1cm}

The rest of this paper is organized as follows. Section~\ref{sec:relatedwork} reviews related work. Section~\ref{sec:dataset} describes the dataset construction process. Section~\ref{sec:securityIssues} through Section~\ref{sec:rejection} respectively present our analyses for addressing RQ1 through RQ4 outlined above. Section~\ref{sec:implications} discusses the implications of our results for researchers, developers, and AI coding agents. Section~\ref{sec:threats} outlines the threats to validity. Finally, Section~\ref{sec:conclusion} concludes the paper and suggests future research directions.

\section{Related Work}
\label{sec:relatedwork}

\subsection{AI in Software Development}

LLM-based coding agents have advanced from code completion systems~\cite{yang2025empirical,wu2025humanevalcomm,sapkota2025vibe} to autonomous agents that can generate and submit PRs~\cite{li2025rise, wang2025ai}. Large scale studies~\cite{li2025rise, watanabe2025use} show that agent generated PRs are now common in OS projects, but they still face lower acceptance rates than human authored ones. Empirical analyses of Claude Code~\cite{watanabe2025use} and Copilot pull requests~\cite{xiao2024generative, nattamai2025impact} reveal that AI contributions are often merged quickly for routine edits but encounter skepticism for security related changes.

Trust and adoption remain central challenges~\cite{roychoudhury2025agentic}. Prior work finds that developers examine AI code more critically because of missing rationale, maintainability concerns, and integration costs~\cite{wang2024investigating, basha2025trust, afroogh2024trust}. Industry reports echo this concern, emphasizing that transparency and traceability are essential for acceptance~\cite{storer2025fostering}.

Existing research primarily examines AI contributions in general contexts, while security focused PRs pose unique challenges. These submissions often involve small but critical modifications where assurance and correctness are essential. To our knowledge, no prior study systematically investigates security-related AI generated PRs. Our work addresses this gap by analyzing their characteristics, review dynamics, and acceptance patterns.

\subsection{Security Contributions in OS Projects}

Prior work shows that security patches differ markedly from general bug fixes. They are usually smaller in scope, involve higher risk, and undergo stricter review~\cite{li2017large, lin2024vulnerabilities, woo2025large, dissanayake2022software}.
PatchDB highlights the fine grained structure of human authored security patches linked to CVEs~\cite{wang2021patchdb}, while other studies show that security fixes often require specialized workflows for identification and validation~\cite{lin2024vulnerabilities}.
Recent work has also examined silent security patches without CVE references~\cite{tang2025just} and the threat of malicious commits that appear benign~\cite{wu2021feasibility}.
These studies provide a detailed understanding of how human developers handle security patches in OS projects. However, no published research has examined security PRs made by coding agents. As a result, it remains unclear whether agent generated security contributions follow the same patterns of size, review rigor, and risk as human authored patches.

\subsection{Pull Request Review Dynamics}

Early studies identify both technical and social factors that shape review outcomes. Project characteristics, code quality, and contributor reputation affect acceptance, while relationships between reviewers and contributors and the tone of discussion also matter~\cite{tsay2014influence, gousios2015work}. Zhang et al.~\cite{zhang2022pull2} show that a small group of variables, such as whether the contributor and integrator are the same person, explain much of the variance in acceptance, although their effects differ with project maturity and community context.
Beyond acceptance, review latency has emerged as a central measure of productivity and contributor engagement. Yu et al.~\cite{yu2015wait} report that contributor experience and project size influence latency. Zhang et al.~\cite{zhang2022pull} further show that description length and code churn affect early delays, while review comments and continuous integration outcomes drive later ones.
Most of this evidence comes from human authored PRs. It is not yet known whether the same review dynamics apply to security-related AI generated PRs, which involve unique technical risks and require higher assurance for trust.

\subsection{Commit Message Quality}

Commit messages play a central role in communicating code changes during pull request reviews. Messages that clearly state both what was changed and why help reviewers understand intent and assess correctness. Prior studies show that many commit logs fall short of this goal. Imtiaz et al.~\cite{tian2022makes} report that about 44\% of messages omit either the “what” or the “why,” and Li et al.~\cite{li2023commit} find that missing rationale increases the chance of defects and reduces the likelihood of acceptance. Li et al.~\cite{li2025optimization} further show that messages with clear semantics improve reviewer understanding and decision outcomes.
Most of this research focuses on human written commits. The quality and influence of messages produced by coding agents remain largely unexplored, leaving open how clarity and rationale affect the acceptance of AI generated PRs.

\subsection{Rejection Reasons and Taxonomies}

Prior research has organized the causes of pull request rejections into structured taxonomies.
Pantiuchina et al.~\cite{pantiuchina2021developers} analyzed review comments on refactoring requests and identified recurring rejection patterns that fall into two broad groups: process-related and refactoring-specific reasons.
Building on this work, Watanabe et al.~\cite{watanabe2025use} examined large collections of agentic PRs and reused the same framework while adding new categories observed in autonomous submissions such as “submission for verification.”
Existing taxonomies, however, are based on general PRs that include refactorings, maintenance updates, and mixed changes, or on general agentic submissions.
They do not reflect the distinct properties of security-focused contributions.
This difference raises the need for a closer examination of how rejection reasons manifest in security-related AI-generated PRs.

\section{Data Collection}
\label{sec:dataset}

\subsection{Source Dataset}

We focus on security-related PRs submitted by agentic AI. 
To construct this dataset, we build upon the recently released AIDev dataset~\cite{li2025rise}, which contains over 456K PRs generated by AI coding agents such as OpenAI Codex~\cite{openai2025codex}, Devin~\cite{cognition2025devin}, GitHub Copilot~\cite{github2025copilot}, Cursor~\cite{cursor2025}, and Claude Code~\cite{anthropic2025claudecode}. The dataset is actively maintained; we use the version last updated on August 1, 2025.

From the full AIDev dataset, we first filter repositories with more than 100 GitHub stars, as such projects are more likely to exhibit sustained community engagement and code review activity.
This filtering step yields 2,807 repositories. From these repositories, we extract all PRs authored by AI agents, resulting in 33,596 AI-generated PRs that form the foundation for our subsequent security-focused analysis.

\begin{figure*}[htbp]
\centering
\includegraphics[scale=0.43]{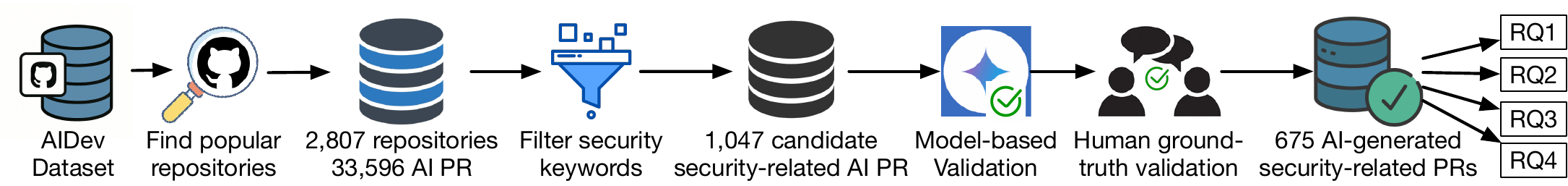}

\caption{Overview of our data construction process}
\label{fig:method}
\end{figure*}

Notably, Devin’s earliest contributions begin on December 12, 2024, and among our collected PRs, 65 AI PRs precede January 1, 2025. 
Figure~\ref{fig:method} illustrates the construction process of our security-related AI PRs dataset derived from the AIDev dataset.

\subsection{Identifying Security Related PRs}

We aim to construct a reliable dataset of security-related PRs from the larger collection of AI generated ones. Keyword-based filtering is a widely used technique in software engineering studies~\cite{peters2017text} and provides a systematic way to identify relevant commits and PRs. To design our filtering process, we combine curated keyword lists from prior work by Paul et al.~\cite{paul2021security}, Zhou et al.~\cite{zhou2021spi}, and Zhou and Sharma~\cite{zhou2017automated}. Together these studies provide more than one hundred terms that capture various vulnerability types, regular-expression patterns, and commit-level indicators of security fixes.

Our consolidated list contains 66 strong security keywords and 11 generic fix terms, as summarized in Table~\ref{tab:security_keywords}. To reduce false positives, we adopt a conservative two-tier filtering strategy. Strong security terms are matched directly, while generic fix terms are counted only when they occur in the same sentence or within 100 characters of a strong security keyword.
This proximity-based condition ensures that general fix phrases are retained only when they clearly express a security-related intent.

\begin{table}[htbp]
\centering
\caption{Keywords used for identifying security-related PRs}
\label{tab:security_keywords}

\begin{tabular}{|p{15cm}|}
\hline
\textbf{Strong Security Keywords} \\ 
\hline
CVE, CWE, vulnerability, vulnerable, exploit, exploitable, security flaw, zero day, one day, hidden vuln, latent vuln, XSS, SQLi, injection, CSRF, RCE, ReDoS, XXE, open redirect, buffer overflow, heap overflow, stack overflow, off by one, format string, path traversal, directory traversal, privilege escalation, unauthorized, bypass, sandbox escape, leak, data leak, info leak, information disclosure, memory disclosure, malicious, backdoor, trojan, spyware, payload, exploit kit, shellcode, hotfix, security patch, vulnerability fix, DoS, denial of service, crash, infinite loop, OOM, unsafe, race, race condition, dangling pointer, use after free, null dereference, memory corruption, concurrent access, time of check time of use, sanitize, sanitization, unvalidated input, improper input validation, tainted input, validate input, parameter tampering \\
\hline \hline
\textbf{Generic Fix Terms (checked for co-occurrence)} \\ 
\hline
fix, fixes, fixed, patch, patched, resolve, resolves, resolved, repair, mitigate, mitigation \\
\hline
\end{tabular}
\end{table}

Applying this filtering process to the 33,596 AI-generated PRs yields 1,047 security-related PRs, representing 3.1\% of the total. These PRs encompass a broad range of security indicators. The most frequent ones include \emph{injection} (162 instances), \emph{crash} (144), \emph{race} (140), and \emph{payload} (101). Other recurring indicators such as \emph{vulnerability} (90), \emph{unsafe} (85), \emph{bypass} (69), \emph{race condition} (67), \emph{leak} (62), and \emph{sanitize} (58). 

\subsection{Validation and Final Dataset}

We perform a model-based validation using the \textit{gemini-2.0-flash} API to automatically verify whether each PR is security related, providing the PR title and body as input.
The model outputs a binary label (Yes/No) with a short rationale. Recent studies show that LLMs can achieve near-human accuracy in annotation and labeling tasks when properly prompted~\cite{sayeed2024annotating, vuruma2025automated, gilardi2023chatgpt}.

This validation step reduces the set to \textbf{675 security-related AI-generated PRs}. To check how well the Gemini model performed, two authors manually label a representative sample from the 675 PRs, following the sampling approach used in prior software-engineering studies~\cite{ghaleb2019studying, kondo2022empirical, macho2021nature, fregnan2022happens}. At a 95\% confidence level with a $\pm$5\% margin of error~\cite{hazra2017using}, the required sample size is approximately 245 PRs. The inter-rater reliability, measured by Cohen’s $\kappa$~\cite{vieira2010cohen}, is 0.79, which indicates substantial agreement between the human annotators. All analyses in the paper use the 675 PRs labeled by the Gemini model; the human-labeled subset is used only to verify annotation reliability. We use this validated dataset~\cite{dataset} in all analyses for RQ1–RQ4.


\begin{table}[htbp]
\centering
\caption{Status distribution of 675 security-related AI PRs}
\label{tab:pr_status}

\begin{tabular}{|l|c|c|}
\hline
\textbf{Status} & \textbf{Count} & \textbf{Percentage} \\
\hline
Merged & 354 & 52.4\% \\\hline
Closed (not merged) & 219 & 32.4\% \\\hline
Open & 102 & 15.1\% \\
\hline
\end{tabular}
\end{table}

Within the complete set of 675 security-related AI-generated PRs, the distribution of outcomes is summarized in Table \ref{tab:pr_status}.
More than half of the PRs (52.4\%) are merged, nearly one-third (32.4\%) are closed without merging, and the remainder (15.1\%) remain open.


\section{Security Issues Introduced by AI PRs (RQ1)} \label{sec:securityIssues}
\subsection{Approach}

To examine the types of security issues introduced by security-related AI-generated PRs, we analyze their code changes using Semgrep~\cite{semgrep} (version 1.135.0). We select Semgrep for four reasons:

\begin{itemize}
    \item It provides advanced static analysis rules that cover a wide range of vulnerability classes and ensure consistent detection across different projects.

    \item Our dataset includes repositories written in several programming languages. Semgrep supports  these languages, such as Java, Python, Go, Rust, JavaScript, TypeScript, PHP, Ruby, and Swift, as well as multiple package managers and manifest files.

    \item It allows full automation and reproducible scans, which makes it suitable for large scale empirical analysis.

    \item It has been adopted in several recent software engineering and security studies, which shows its reliability and community acceptance~\cite{bennett2024semgrep, kluban2024detecting, li2024evaluating, munson2025little}.
\end{itemize}

For each security-related AI PR, we perform Semgrep scans on two code versions: the \textit{baseline commit}, which represents the state of the repository before the PR, and the \textit{PR commit}, which contains the proposed changes. This setup ensures that only new security issues introduced by the PR are detected.

For every repository, we clone the project and fetch the branch containing the PR. The corresponding baseline commit is identified as the version that exists just before the PR branch diverges from the default branch, such as \texttt{main} or \texttt{master}. Semgrep scans are executed on both versions, and the results are exported as JSON files. We then compare these outputs to identify vulnerabilities that appear only in the PR version.

This procedure is repeated for all repositories in our dataset of security-related AI-generated PRs. The resulting dataset contains structured information about newly introduced security issues, including rule identifiers, issue types, file paths, and affected code regions. We use this dataset to analyze the types of vulnerabilities that AI agents introduce.
\subsection{Results}

We apply Semgrep across all 675 security-related AI generated PRs to detect newly introduced vulnerabilities. Out of the full set, 104 PRs (15.4\%) introduce at least one issue, while the remaining 571 PRs (84.6\%) show no Semgrep alerts. In total, Semgrep reports 853 findings. Among the affected PRs, the median number of issues per PR is three, while the maximum reaches 109 in a single PR. The PR with the highest number of alerts is rejected, which suggests that contributions with unusually high concentrations of security alerts are less likely to pass review.

\paragraph{Distribution of Vulnerability Types.} Table~\ref{tab:topcwes} presents the most frequent CWE categories detected by Semgrep. A few categories dominate the results. Inefficient regular expression complexity (CWE-1333) accounts for 36.2\% of all findings, followed by OS (operating system) command injection (CWE-78, 13.0\%) and path traversal (CWE-22, 10.3\%). Other recurring types include use of externally controlled format strings (CWE-134, 8.2\%), cross site scripting (CWE-79, 7.1\%), and use of hard coded credentials (CWE-798, 5.7\%). Together, these six categories represent more than 80\% of all detected weaknesses.

\begin{table}[htbp]
\centering
\caption{Top CWE types in security-related AI PRs}
\label{tab:topcwes}

\begin{tabular}{|l@{}|l|@{}r|@{}r|}
\hline
\textbf{CWE ID} & \textbf{CWE Name} & \textbf{Count} & \textbf{Percent} \\
\hline \hline
CWE-1333 & Inefficient Regular Expression Complexity & 306 & 36.2\% \\\hline
CWE-78   & OS Command Injection                      & 110 & 13.0\% \\\hline
CWE-22   & Path Traversal                            & 87  & 10.3\% \\\hline
CWE-134  & Use of Externally-Controlled Format String               & 69  & 8.2\% \\\hline
CWE-79   & Cross-Site Scripting (XSS)                & 60  & 7.1\% \\\hline
CWE-798  & Use of Hard-coded Credentials                    & 48  & 5.7\% \\\hline
CWE-89   & SQL Injection                             & 17  & 2.0\% \\\hline
CWE-116  & Improper Encoding or Escaping of Output    & 16  & 1.9\% \\\hline
CWE-319  & Cleartext Transmission of Sensitive Data  & 14  & 1.6\% \\\hline
CWE-470  & Unsafe Reflection                         & 14  & 1.6\% \\\hline
--       & Others                          & 112 & 13.2\% \\\hline
\end{tabular}
\end{table}

The dominance of CWE-1333 suggests that AI systems often generate regular expressions without verifying efficiency or safety, which can lead to denial of service risks. Similarly, the frequent presence of injection flaws such as command injection (CWE-78) and SQL injection (CWE-89) indicates that AI agents sometimes form shell commands or database queries through unsafe string concatenation. Path traversal (CWE-22) and format string issues (CWE-134) also appear often, suggesting weak input validation and limited secure coding awareness.

The long tail of less frequent CWEs (grouped as “Others,” 13.2\%) includes 11 distinct categories, each with 10 or fewer findings. These cover cryptographic issues, open redirects, deserialization flaws, improper authorization, and other isolated weaknesses. The diversity of issues indicates that AI PRs touch on nearly every major class of security weakness, even though a small subset of categories dominates the overall distribution.



\paragraph{Variation Across AI PR Categories and Outcomes.}

We further analyze vulnerabilities across the five categories of AI-generated PRs and their review outcomes (\textit{accepted}, \textit{rejected}, or \textit{open}). Table~\ref{tab:ai_pr_outcome} summarizes the distribution of Semgrep detected security issues across agents and outcomes. For each agent, it reports the number of PRs by outcome, the percentage of PRs containing at least one detected issue, and the top three CWE types identified in each group along with their counts.

\begin{table}[h]
\centering
\caption{Semgrep issues by AI PR category and PR outcome}
\label{tab:ai_pr_outcome}

\begin{tabular}{|l|l|r|r|p{7cm}|}
\hline
\textbf{Cate-} & \textbf{PR Out} & \textbf{Total} & \textbf{\% with} & \textbf{Top 3 CWEs (count)} \\
\textbf{gory} & \textbf{-come} & \textbf{PRs} & \textbf{$\geq$1 Issue} &\\
\hline \hline
  & Accepted & 22  &  22.7\% & CWE-532 (6), CWE-134 (5), CWE-918 (3) \\\cline{2-5}
             Claude & Open     & 2   & 0.0\%  & -- \\\cline{2-5}
              Code & Rejected & 9   & 11.1\% & CWE-78 (2) \\\hline
       & Accepted & 110 & 11.8\% & CWE-22 (12), CWE-79 (9), CWE-78 (5) \\\cline{2-5}
            Copilot  & Open     & 69  &21.7\% & CWE-79 (14), CWE-22 (9), CWE-639 (8) \\\cline{2-5}
              & Rejected & 103 &14.6\% & CWE-1333 (289), CWE-78 (92), CWE-798 (40) \\\hline
       & Accepted & 27  &11.1\% & CWE-134 (4), CWE-1333 (3), CWE-327 (1) \\\cline{2-5}
          Cursor     & Open     & 12  &41.7\% & CWE-134 (19), CWE-295 (5), CWE-319 (1) \\\cline{2-5}
              & Rejected & 15  &20.0\% & CWE-319 (3), CWE-134 (2), CWE-798 (1) \\\hline
       & Accepted & 74  &20.3\% & CWE-22 (16), CWE-79 (9), CWE-1333 (7) \\\cline{2-5}
        Devin        & Open     & 14  &21.4\% & CWE-78 (3), CWE-134 (2), CWE-116 (1) \\\cline{2-5}
              & Rejected & 76  &13.2\% & CWE-134 (35), CWE-22 (13), CWE-1333 (6) \\\hline
 & Accepted & 121 &12.4\% & CWE-22 (29), CWE-470 (14), CWE-95 (8) \\\cline{2-5}
          OpenAI   & Open     & 5   &20.0\% & CWE-79 (2), CWE-522 (1) \\\cline{2-5}
          Codex      & Rejected & 16  &0.0\%  & -- \\\hline
\end{tabular}

\end{table}

We observe distinct patterns across PR sources and review outcomes. Copilot, which contributes the largest number of PRs (282), shows issues across all outcome groups.  Rejected Copilot PRs display the highest concentration of alerts, with 15 PRs (14.6\%) flagged and dominated by CWE-1333 (289 instances), CWE-78 (92), and CWE-798 (40). This pattern shows that Copilot PRs containing regex inefficiencies, injection flaws, or exposed credentials are more frequently rejected.  Accepted Copilot PRs show fewer issues, primarily path traversal and cross site scripting.

Cursor exhibits a higher rate of open PRs with issues (41.7\%) compared to accepted (11.1\%) or rejected (20.0\%). The top weaknesses in open Cursor PRs include format string vulnerabilities (CWE-134) and certificate validation errors (CWE-295). This outcome pattern shows that Cursor PRs with detected security issues are more likely to remain open than be accepted or rejected.

Devin PRs show a relatively even spread across outcomes. Accepted Devin PRs frequently introduce path traversal and cross-site scripting, whereas rejected ones contain many format string problems (CWE-134, 35 instances). This recurrence suggests an association between format string misuse and rejection outcomes.
Claude Code contributes fewer PRs overall but still introduces issues such as sensitive data exposure in log files (CWE-532) and server side request forgery (CWE-918). Interestingly, even with a smaller scale, accepted Claude PRs contain issues, which highlights that some vulnerabilities pass review despite being detectable by Semgrep.

OpenAI Codex contributions show the lowest proportion of PRs with detected issues. Accepted Codex PRs still include vulnerabilities, particularly path traversal (29 instances) and unsafe reflection (14), whereas no issues appear in its rejected PRs. 
This may suggest that rejection decisions are not always associated with Semgrep-detected flaws. We answer RQ1 as follows:

\vspace{-0.2cm}
\begin{tcolorbox}
[colback=gray!10,colframe=black,boxsep=1pt,left=2pt,right=2pt,top=2pt,bottom=2pt,boxrule=0.2pt] 
\textbf{Ans. to RQ1:} Security-related AI PRs tend to introduce a small set of weaknesses. Most issues fall into regex inefficiency (CWE-1333), injection risks (CWE-78, CWE-89, CWE-94), and path traversal (CWE-22). These dominate our findings, while the remaining CWE types appear only rarely.
We also find differences across agents and review outcomes. Copilot PRs often include regex or injection problems that lead to rejection, Cursor PRs with similar flaws more often remain open, and Devin PRs frequently include format-string misuse that tends to be rejected. A few vulnerabilities still pass through review even in security-focused submissions.

\end{tcolorbox}



\section{Review Latency and Acceptance (RQ2)}\label{sec:latencyAcceptance}
\subsection{Approach}
    
To study the factors that influence PR latency and acceptance in security-related AI-generated PRs, we build on prior empirical research on pull request processes~\cite{zhang2022pull, yu2015wait, dey2020effect, zhang2022pull2}. We consolidate factors examined in these works and select those that are both theoretically meaningful and practically extractable from our dataset.
We extract 24 factors covering requester experience, PR content, social interaction, project characteristics, and testing signals. The data snapshot is taken on September 15, 2025. Table~\ref{tab:factors} summarizes each factor, and its rationale for inclusion.

\begin{table*}[t]
\centering
\small
\caption{Summary of 24 factors considered for PR latency and acceptance analysis}
\label{tab:factors}

\setlength\tabcolsep{1.5pt}
\begin{tabular}{|p{2.5cm}|p{3.45cm}|p{9.8cm}|}
\hline
\textbf{Factor} & \textbf{Description} & \textbf{Rationale} \\
\hline
first\_pr & Whether the PR is the \newline requester’s first in the project & New contributors may face longer reviews and lower acceptance due to unfamiliarity with project conventions and limited prior trust~\cite{zhang2022pull, zhang2022pull2}. \\
\hline
prior\_review\_num & \# of prior reviews performed by the requester & Requesters who have reviewed more PRs in the project are familiar with its practices, which may lead to faster responses and higher acceptance rates~\cite{zhang2022pull, yu2015wait, zhang2022pull2}. \\
\hline
requester\_succ\_rate & Historical PR acceptance rate of the requester in the project & A requester’s historical success rate reflects their reputation and credibility, which may increase reviewer confidence and merge likelihood~\cite{zhang2022pull, yu2015wait, dey2020effect, zhang2022pull2}. \\
\hline
num\_hash\_tag & \# of hashtags in PR description & Hashtags link PRs to issues or milestones, providing context that may support faster reviews~\cite{zhang2022pull, yu2015wait, zhang2022pull2}. \\
\hline
num\_at\_mentions & \# of user mentions in PR \newline description & Mentioning specific users may draw direct attention from maintainers or reviewers, which may reduce review latency. \\
\hline
description\_length & Length (in \# of characters) of PR description & A detailed description may signal preparation and seriousness, but also higher complexity that can extend review time~\cite{zhang2022pull,yu2015wait}. \\
\hline
friday\_effect & PR opened on a Friday & PRs opened near weekends often face slower review cycles due to reduced reviewer activity~\cite{zhang2022pull, yu2015wait, zhang2022pull2}. \\
\hline
project\_age \newline (minutes) & Age of project at PR creation & Mature projects tend to have more established workflows and norms, which can influence both review speed and acceptance~\cite{zhang2022pull, yu2015wait, zhang2022pull2}. \\
\hline
stars & \# of GitHub stars of project & Highly starred repositories attract broader community attention and often undergo stricter reviews~\cite{zhang2022pull2}. \\
\hline
forks & \# of forks of project & Projects with many forks show high engagement, which may boost activity but add review backlogs~\cite{zhang2022pull2}. \\
\hline
open\_pr\_num & \# of open PRs in the project & Many concurrent open PRs may indicate review congestion, leading to slower responses~\cite{zhang2022pull, yu2015wait, zhang2022pull2}. \\
\hline
num\_comments & \# of review comments in PR thread & More comments may signal active discussion or disagreement, which may lead to longer review times~\cite{zhang2022pull, yu2015wait, dey2020effect, zhang2022pull2}. \\
\hline
num\_participants & \# of unique participants in PR thread & A larger number of participants may bring diverse feedback that may accelerate review decisions or increase scrutiny~\cite{zhang2022pull}. \\
\hline
commits & \# of commits in PR & PRs with multiple commits may be more complex and may require additional verification effort~\cite{yu2015wait,dey2020effect}. \\
\hline
additions & Lines of code added by PR & Large additions may expand code surface, potentially increasing review time and rejection risk~\cite{zhang2022pull,dey2020effect}. \\
\hline
deletions & Lines of code removed by PR & Large deletions may simplify maintenance but raise concern if critical functionality is removed~\cite{zhang2022pull,dey2020effect}. \\
\hline
changed\_files & \# of files changed by PR & PRs modifying many files may have broader scope and need review from multiple experts~\cite{zhang2022pull,dey2020effect,zhang2022pull2}. \\
\hline
pr\_succ\_rate & Historical PR acceptance rate of project & The project’s overall merge rate may reflect its governance style and openness to external contributions~\cite{zhang2022pull2}. \\
\hline
ci\_exists & Whether continuous \newline integration is configured & Projects with CI configured may provide automated feedback, which may build reviewer trust and reduce delays~\cite{zhang2022pull,yu2015wait,zhang2022pull2}. \\
\hline
ci\_latency (minutes) & Time taken for CI checks & Longer CI may delay PR review~\cite{zhang2022pull,yu2015wait,zhang2022pull2}. \\
\hline
sloc & Source lines of code in project & Larger projects may have longer review times~\cite{zhang2022pull,zhang2022pull2}. \\
\hline
test\_cases\_per\_kloc & \# of test cases per 1k LOC & Higher test density may reflect stronger testing practices and increase reviewer confidence~\cite{zhang2022pull,zhang2022pull2}. \\
\hline
test\_lines\_per\_kloc & \# of test LOC per 1k LOC & Greater test code coverage may indicate better quality and raise merge likelihood~\cite{zhang2022pull,zhang2022pull2}. \\
\hline
contain\_test\_code & Whether PR modifies test files & PRs with test changes may appear safer and more complete, improving acceptance chances~\cite{yu2015wait,dey2020effect,zhang2022pull2}. \\
\hline
\end{tabular}
\end{table*}

For both PR latency and acceptance analyses, we apply the consistent preprocessing techniques to ensure data quality and model interpretability. We handle missing values using \textit{k}-nearest neighbor (KNN) imputation~\cite{zhang2012nearest} with $k=5$.   
We then address infinite values, which occur in derived ratio-based features such as \texttt{requester\_succ\_rate}. Infinite values can destabilize regression coefficients and statistical tests~\cite{osborne2004power, harrell2015multivariable}. We therefore replace them with zero, a conservative strategy that prevents invalid computations while keeping the feature in the analysis.  

To reduce the influence of outliers and improve the linearity assumption of regression models, we apply a logarithmic transformation~\cite{benoit2011linear} to skewed numeric features, a common strategy in regression modeling~\cite{harrell2015multivariable}. This transformation compresses long right tails of distributions (e.g., \texttt{additions} and \texttt{deletions}) and places variables with wide dynamic ranges on a comparable scale. Binary variables (e.g., \texttt{first\_pr}, \texttt{friday\_effect}, \texttt{ci\_exists}) remain unchanged since they do not require transformation.  

We mitigate multicollinearity among factors using the \texttt{AutoSpearman} algorithm~\cite{okoye2024correlation}. Multicollinearity inflates variance in coefficient estimates and reduces the interpretability of regression results. \texttt{AutoSpearman} automatically detects pairs of variables with Spearman correlation $\rho \geq 0.7$ and iteratively removes one variable from each pair, retaining the more informative factor.

After preprocessing, we perform two separate regression analyses. 
For \emph{PR latency}, we fit a linear regression model on the log-transformed response time~\cite{benoit2011linear}. 
For \emph{PR acceptance}, we fit a logistic regression model~\cite{gey1994inferring} that estimates the probability of a PR being merged. 
Both models use the reduced feature set produced by \texttt{AutoSpearman}.

We report evaluation measures aligned with each analysis.
For PR latency, we present regression coefficients, their statistical significance, and the model’s $R^{2}$ and adjusted $R^{2}$ values~\cite{harrell2015multivariable}.
For PR acceptance, we report odds ratios (OR), 95\% confidence intervals, and corresponding $p$-values to assess the strength and direction of effects~\cite{hosmer2013applied}.
For both models, we interpret significant factors and the overall explanatory power based on their $R^{2}$ or pseudo-$R^{2}$, following established reporting guidelines in empirical software engineering~\cite{kitchenham2022segress}.
\subsection{Results}

\paragraph{PR Latency.}
Table~\ref{tab:latency} presents the linear regression results for PR latency. The model achieves an adjusted $R^{2}$ of 0.574, which means that the included factors explain a substantial proportion of the observed variance. 

\begin{table}[htbp]
\centering
\caption{Linear regression results for PR latency}
\label{tab:latency}

\begin{tabular}{|l|r|r|r|}
\hline
\textbf{Factor} & \textbf{Estimate} & \textbf{Std. Error} & \textbf{$p$-value} \\
\hline \hline

log\_prior\_review\_num & -0.259 & 0.062 & $<0.001$ *** \\\hline
log\_num\_hash\_tag     &  0.263 & 0.100 & 0.009 ** \\\hline
log\_num\_at\_mentions  &  0.011 & 0.196 & 0.957 \\\hline
log\_project\_age       & -0.072 & 0.079 & 0.363 \\\hline
log\_stars              &  0.135 & 0.052 & 0.010 * \\\hline
log\_open\_pr\_num      &  0.005 & 0.066 & 0.938 \\\hline
log\_num\_comments      &  0.295 & 0.129 & 0.023 * \\\hline
log\_commits            & -0.008 & 0.191 & 0.968 \\\hline
log\_additions          &  0.087 & 0.052 & 0.092  \\\hline
log\_deletions          & -0.004 & 0.056 & 0.945 \\\hline
log\_pr\_succ\_rate     & -5.339 & 1.077 & $<0.001$ *** \\\hline
log\_ci\_latency        &  0.622 & 0.045 & $<0.001$ *** \\\hline
log\_sloc               & -0.050 & 0.064 & 0.436 \\\hline
log\_test\_lines\_per\_kloc & 0.000 & 0.048 & 0.995 \\\hline
requester\_succ\_rate   & -1.518 & 0.323 & $<0.001$ *** \\\hline
first\_pr               & -1.813 & 0.328 & $<0.001$ *** \\\hline
friday\_effect          &  0.012 & 0.246 & 0.962 \\\hline
ci\_exists              & -1.958 & 0.298 & $<0.001$ *** \\\hline
contain\_test\_code     &  0.538 & 0.214 & 0.012 * \\\hline
\multicolumn{4}{c}{Here, significance codes: * $p<0.05$, ** $p<0.01$, *** $p<0.001$.}
\end{tabular}
\end{table}

Prior review activity is an important determinant. The number of prior reviews performed by the requester (\texttt{log\_prior\_review\_num}) is negatively associated with latency ($p<0.001$). This means that contributors with more review experience tend to receive faster evaluations. Social signals also play a role. A higher number of hashtags in the PR description increases review time ($p<0.01$), and PRs that draw more reviewer comments also take longer to close ($p<0.05$). These findings suggest that socially complex or debated contributions extend the review cycle.

Project-level characteristics further influence latency. PRs in projects with more stars experience longer review times ($p<0.05$). Popular repositories likely receive heavier traffic and stricter review. In contrast, the project’s prior PR success rate is linked to shorter latency ($p<0.001$). Projects that merge contributions more often also review them more quickly.

CI-related variables show strong effects. Longer CI runtime (\texttt{log\_ci\_latency}) increases latency ($p<0.001$). At the same time, the existence of CI itself (\texttt{ci\_exists}) reduces latency ($p<0.001$). These results highlight the efficiency benefits of integrated workflows. Requester experience is also central. A higher requester success rate reduces latency ($p<0.001$), and PRs from first-time contributors are processed faster ($p<0.001$). Finally, PRs that contain test code take longer to review ($p<0.05$). This may indicate that reviewers invest additional time in validating such changes.
Other features do not show significant effects. 




\paragraph{PR Acceptance.}

Table~\ref{tab:acceptance} presents the logistic regression results for PR acceptance. The model has a pseudo-$R^{2}$ of 0.23, which gives moderate explanatory power. 

\begin{table}[htbp]
\centering
\caption{Logistic regression results for PR acceptance}
\label{tab:acceptance}

\begin{tabular}{|l|r|r|r|}
\hline
\textbf{Factor} & \textbf{Estimate} & \textbf{OR} & \textbf{$p$-value} \\
\hline \hline

log\_prior\_review\_num &  0.188 & 1.21 & 0.013 * \\\hline
log\_num\_hash\_tag     & -0.331 & 0.72 & 0.006 ** \\\hline
log\_num\_at\_mentions  & -0.319 & 0.73 & 0.150 \\\hline
log\_project\_age       & -0.052 & 0.95 & 0.568 \\\hline
log\_stars              & -0.067 & 0.94 & 0.266 \\\hline
log\_open\_pr\_num      & -0.197 & 0.82 & 0.011 * \\\hline
log\_num\_comments      &  0.433 & 1.54 & 0.005 ** \\\hline
log\_commits            & -0.087 & 0.92 & 0.693 \\\hline
log\_additions          & -0.040 & 0.96 & 0.528 \\\hline
log\_deletions          &  0.065 & 1.07 & 0.326 \\\hline
log\_pr\_succ\_rate     &  5.636 & 280.26 & $<0.001$ *** \\\hline
log\_ci\_latency        &  0.178 & 1.19 & 0.001 ** \\\hline
log\_sloc               &  0.027 & 1.03 & 0.719 \\\hline
log\_test\_lines\_per\_kloc & -0.012 & 0.99 & 0.838 \\\hline
requester\_succ\_rate   &  2.315 & 10.13 & $<0.001$ *** \\\hline
first\_pr               &  1.451 & 4.27 & $<0.001$ *** \\\hline
friday\_effect          &  0.028 & 1.03 & 0.921 \\\hline
ci\_exists              &  0.041 & 1.04 & 0.908 \\\hline
contain\_test\_code     & -0.673 & 0.51 & 0.006 ** \\\hline
\multicolumn{4}{c}{Here, significance codes: * $p<0.05$, ** $p<0.01$, *** $p<0.001$}
\end{tabular}
\end{table}

Requester-related variables have the strongest influence. The requester success rate is highly predictive of acceptance (OR = 10.13, $p<0.001$). PRs from requesters with a stronger history of merged contributions are far more likely to be accepted. First-time contributors also face higher odds of merging (OR = 4.27, $p<0.001$). In addition, prior review activity of the requester is positively associated with acceptance (OR = 1.21, $p<0.05$), which indicates that active reviewers in the community gain credibility when they later submit PRs.

Project-level factors also show clear effects. Projects with a higher overall PR success rate are much more likely to merge new AI security PRs (OR = 280.26, $p<0.001$). In contrast, projects with many open PRs show a lower chance of merging new submissions (OR = 0.82, $p<0.05$). This pattern indicates that established projects with consistent merging practices handle AI contributions more efficiently, while overloaded projects merge fewer PRs overall.

Social interaction factors contribute as well. The number of reviewer comments is positively associated with acceptance (OR = 1.54, $p<0.01$). This suggests that discussion and reviewer engagement can improve the quality of a PR or increase reviewer confidence. On the other hand, descriptive complexity works in the opposite direction. PR descriptions with more hashtags are less likely to be accepted (OR = 0.72, $p<0.01$).

CI-related signals influence outcomes. PRs with longer CI runtimes are more likely to be merged (OR = 1.19, $p<0.01$). This may indicate that reviewers trust changes that survive more extensive testing. Finally, test-related evidence reduces acceptance. The inclusion of test code is negatively associated with merging (OR = 0.51, $p<0.01$). This unexpected effect suggests that AI-generated tests may be noisy or fail to provide convincing evidence of correctness. Based on these findings, we answer RQ2 as follows.

\vspace{-0.2cm}
\begin{tcolorbox}[colback=gray!10,colframe=black,boxsep=1pt,left=2pt,right=2pt,top=2pt,bottom=2pt,boxrule=0.2pt] 

\textbf{Ans. to RQ2:} Review latency decreases when the requester has more prior reviews or a higher past success rate, and when the project has a higher merge rate. CI presence and first-time submissions also reduce latency, while long CI runtimes slow the review. On the other hand, acceptance depends on similar factors. Projects with higher merge rates and requesters with stronger prior merge records see higher acceptance, and first-time contributors also have a higher chance of getting their security-related AI PRs merged.
\end{tcolorbox}



\section{Commit Message Quality (RQ3)}\label{sec:msgQuality}
\subsection{Approach}

Commit messages are most informative when they convey both the \emph{What} (a summary of the change) and the \emph{Why} (its rationale)~\cite{tian2022makes}. We adopt the \emph{C-Good} model introduced by Tian et al. \cite{tian2022makes}, which classifies whether a commit message contains these two elements.
Li and Ahmed \cite{li2023commit} later used this model to study temporal trends in message quality.
Following these studies, we reuse the same architecture and preprocessing pipeline to evaluate the commit-message quality of AI-generated security PRs.

We collect commit messages for all 675 PRs using the GitHub API and retain only English messages.
Each message is preprocessed by replacing URLs, version numbers, and code snippets with placeholders; removing sign-off lines; performing tokenization and part-of-speech tagging; and substituting standardized tokens (e.g., \texttt{$url$}, \texttt{$methodName$}).
Messages that become empty after preprocessing are labeled as \textit{empty log messages}.
This pipeline ensures full compatibility with the data format used to train the original model.

The \emph{C-Good} dataset defines four message categories—containing both Why and What, containing neither, missing What, and missing Why—but merges the last three into a single bad class for binary classification.
Thus, messages including both elements are labeled good (0), while others are labeled bad (1).
We retrain the C-Good model on its original annotated dataset using the same hyperparameters.
The model, which combines a BERT encoder with a BiLSTM classifier, achieved a reported precision of 81.6\% in \cite{tian2022makes}.
After retraining, we apply it to our corpus of security-related AI-generated PRs to classify commit-message quality automatically.

To assess reliability in our context, we manually verify a random sample of 339 commit messages (95\% confidence, ±5\% margin~\cite{hazra2017using}).
One annotator labels whether each message includes both What and Why.
The classifier achieves 93.2\% accuracy (precision = 93.8\%, recall = 84.3\%, F1 = 88.8\%, $\kappa$ = 0.84), indicating substantial agreement with human judgment.
These results confirm that the C-Good classifier performs reliably for large-scale analysis of commit-message quality in security-related AI-generated PRs.

\subsection{Results}

We analyze the quality of commit messages across 675 security-related AI-generated PRs, which together include 2,823 commit messages. The analysis is conducted at the commit-message level. We define a high-quality message as one that contains both \textit{Why} and \textit{What} elements, while low-quality messages omit one or both.  
Out of all 2,823 commit messages, 1,988 (70.4\%) are classified as high-quality and 835 (29.6\%) as low-quality. This shows that AI tools often generate messages with both rationale and description.

Table~\ref{tab:cm_quality_agent} shows the distribution of commit message (CM) quality and corresponding PR outcomes for different AI agents. The results vary across tools. Copilot and Devin contribute the largest number of commit messages (911 and 723 high quality, respectively), but their proportions differ. Devin shows a higher share of high quality messages (79.5\%), while Copilot has a lower proportion (71.5\%). Claude Code and Cursor are near the average, with about two thirds of their messages rated as high quality. In contrast, OpenAI Codex produces the lowest share of high quality messages, only 31.3\%, while most (68.7\%) are low quality. These results show that commit message quality differs notably across AI systems.

\begin{table}[htbp]
\centering
\caption{Commit message (CM) quality and PR outcomes}
\label{tab:cm_quality_agent}

\begin{tabular}{|l|l|r|r|r|r|r|}
\hline
\textbf{Agent} & \textbf{Quality} & \textbf{Total} & \textbf{Accepted} & \textbf{Open} & \textbf{Rejected} & \textbf{Acceptance} \\
& & \textbf{CMs} & & & & \textbf{Rate (\%)} \\
\hline \hline
Claude   & High & 188 & 112 & 33 & 43 & 59.6\% \\\cline{2-7}
  Code              & Low  &  83 &  50 & 23 & 10 & 60.2\% \\\hline
Copilot        & High & 911 & 415 & 256 & 240 & 45.6\% \\\cline{2-7}
               & Low  & 364 & 175 &  99 &  90 & 48.1\% \\\hline
Cursor         & High &  95 &  45 &  19 &  31 & 47.4\% \\\cline{2-7}
               & Low  &  45 &  20 &  15 &  10 & 44.4\% \\\hline
Devin          & High & 723 & 267 &  43 & 413 & 36.9\% \\\cline{2-7}
               & Low  & 187 & 101 &  32 &  54 & 54.0\% \\\hline
OpenAI   & High &  71 &  68 &   2 &   1 & 95.8\% \\\cline{2-7}
   Codex            & Low  & 156 & 138 &   5 &  13 & 88.5\% \\\hline
\end{tabular}

\end{table}

We then examine how commit message quality relates to PR outcomes. When messages contain both Why and What, 45.6\% of the corresponding PRs are accepted. When messages are of lower quality, the acceptance rate increases slightly to 58.0\%. This shows that message quality alone may not determine merge decisions, as reviewers likely focus more on the technical correctness of the patch or the reputation of the contributing agent.

Table~\ref{tab:cm_quality_agent} provides a detailed view across AI agents. For Claude Code and Copilot, acceptance rates are similar between high and low quality messages. Devin shows the opposite trend, where PRs with low quality messages are accepted more often. OpenAI Codex shows the clearest benefit from high quality messages, with a noticeable increase in acceptance. Cursor shows little difference between the two groups. These results suggest that the effect of message quality varies across agents and review contexts.

We also compare review latency. PRs with high quality messages close in a mean of 4.31 days (median 0.90), while those with low quality messages close in a mean of 4.46 days (median 1.48). The small difference indicates that message quality does not strongly affect review time, although the higher median latency for low quality messages suggests reviewers may spend more time clarifying or verifying those changes. Based on these findings, we answer RQ3 as follows:

\vspace{-0.2cm}
\begin{tcolorbox}[colback=gray!10,colframe=black,left=2pt,right=2pt,top=2pt,bottom=2pt,boxrule=0.2pt]
\textbf{Ans. to RQ3:} Commit message quality differs across AI agents, and higher quality does not consistently increase acceptance rates or shorten review times.
\end{tcolorbox}

\section{Rejection Reasons (RQ4)}\label{sec:rejection}
\subsection{Approach}
    
\label{sec:rq4_method}

To address this research question, we examine why security-related AI PRs are rejected by project maintainers. Following prior work by Watanabe et al.~\cite{watanabe2025use}, we define a \textit{rejected PR} as one that is closed without being merged. Our objective is to identify and categorize the underlying reasons for these rejections.

We use the rejection taxonomy introduced by Pantiuchina et al. \cite{pantiuchina2021developers} and extended by Watanabe et al. \cite{watanabe2025use}, which captures both technical and process-related reasons for rejection across diverse projects. Although the original taxonomy includes 14 categories, we identify 12 of them in our dataset. 
To cover additional recurring patterns, we apply an \textit{open card sorting approach}~\cite{fincher2005making} in which two authors independently group similar maintainer comments and then discuss differences to reach consensus. 
This process yields \textbf{two new categories}: \textit{Code style or formatting} and \textit{Test failure or insufficient coverage}, which better represent issues specific to AI generated security PRs.

Two authors independently label all rejected PRs using the resulting 14 category taxonomy. 
The inter rater reliability, measured by Cohen’s $\kappa$ = 0.94, indicates \textit{almost perfect agreement}~\cite{landis1977measurement}. 
Disagreements are resolved through discussion until full consensus is achieved. 
In total, we analyze 219 rejected PRs (32.4\% of all AI security-related generated PRs) using this taxonomy.

\subsection{Results}


Table~\ref{tab:rejection_reasons} summarizes the distribution of rejection reasons. The most frequent category is \textit{Unknown} (38.8\%), where PRs are closed without explanatory feedback. This shows that many AI generated contributions are dismissed without maintainers recording their rationale, which limits transparency and learning from rejected work. The second most frequent category is \textit{Are inactive} (12.3\%), where PRs are closed automatically after a period of inactivity rather than through explicit human review. For example, one closure message states, \texttt{"This PR was closed because it has been inactive for 7 days since being marked as stale."} Such automated actions reduce the opportunity for discussion and reflect process level management rather than technical evaluation.

\begin{table}[htbp]

\centering
\caption{Reasons for rejection of AI-generated security PRs}
\label{tab:rejection_reasons}

\small
\begin{tabular}{|p{3cm}|p{6.3cm}|r|}
\hline
\textbf{Category} & \textbf{Description} & \textbf{PRs} \\
\hline \hline
Unknown (No feedback) & PR closed without any reviewer explanation or comment & 38.8\% \\\hline
Are inactive & PR closed after a period of inactivity & 12.3\% \\\hline
Introduce bugs/break APIs & Introduces new defects or breaks backward compatibility & 10.5\% \\\hline
Non-optimal design & PR uses inefficient or poor design choices & 5.9\% \\\hline
Do not add value & Change offers no meaningful improvement to the project & 5.5\% \\\hline
Implemented by others & Similar solution already merged elsewhere & 5.0\% \\\hline
Not sure & Review comments ambiguous; reason unclear & 5.0\% \\\hline
Test failure/lack coverage* & PR fails automated tests or lacks adequate test coverage & 4.1\% \\\hline
Code style / formatting* & Rejected due to style or linting issues & 3.7\% \\\hline
Submission for verification & Opened only to trigger CI pipelines & 3.2\% \\\hline
Not in community interest & Change conflicts with project goals or direction & 2.3\% \\\hline
Distrust in AI-written code & Explicit lack of confidence of AI-generated code & 1.8\% \\\hline
Are obsolete & PR outdated due to new changes & 1.4\% \\\hline
Merge conflicts & PR cannot be merged due to unresolved code conflicts & 0.5\% \\\hline
\multicolumn{3}{r}{
{\small * denotes rejection reasons proposed by this study.}
}
\end{tabular}

\end{table}


Technical issues also appear frequently. \textit{Introduce bugs or break APIs} accounts for 10.5\% of rejections, while \textit{Non optimal design} contributes 5.9\%. These categories highlight that maintainers continue to emphasize correctness and design quality when assessing AI generated PRs. 
Two additional categories appear in this study: \textit{Test failure or insufficient coverage} (4.1\%) and \textit{Code style or formatting} (3.7\%). These cases show that AI PRs are sometimes rejected for specific quality issues such as missing test verification or not meeting linting rules. In contrast, some categories from prior work~\cite{watanabe2025use, pantiuchina2021developers}, such as \textit{Are too large} and \textit{Increase complexity}, do not appear in our dataset. This suggests that the size or complexity issues reported in general AI PR studies are less common in security-focused AI PRs.

Figure~\ref{fig:rejection_heatmap} shows how rejection reasons vary across AI agents. Copilot has the highest number of rejected PRs, many falling under \textit{Introduce bugs or break APIs} (10.7\%) and \textit{Do not add value} (8.7\%). 
Claude shows a greater share of \textit{Implemented by others} (22.2\%) and \textit{Unknown} (33.3\%), reflecting limited reviewer feedback. 
Devin exhibits the most \textit{Are inactive} rejections (31.6\%), consistent with automated closure behavior. 
Cursor has nearly half of its rejected PRs labeled as \textit{Introduce bugs or break APIs} (46.7\%), suggesting recurring functional issues. 
Codex shows the largest proportion of rejections without feedback (56.2\%), indicating that its PRs are often closed automatically or without explicit maintainer commentary. Based on these results, we answer RQ4 as follows:

\begin{figure}[htbp]
\centering
\includegraphics[scale=0.42]{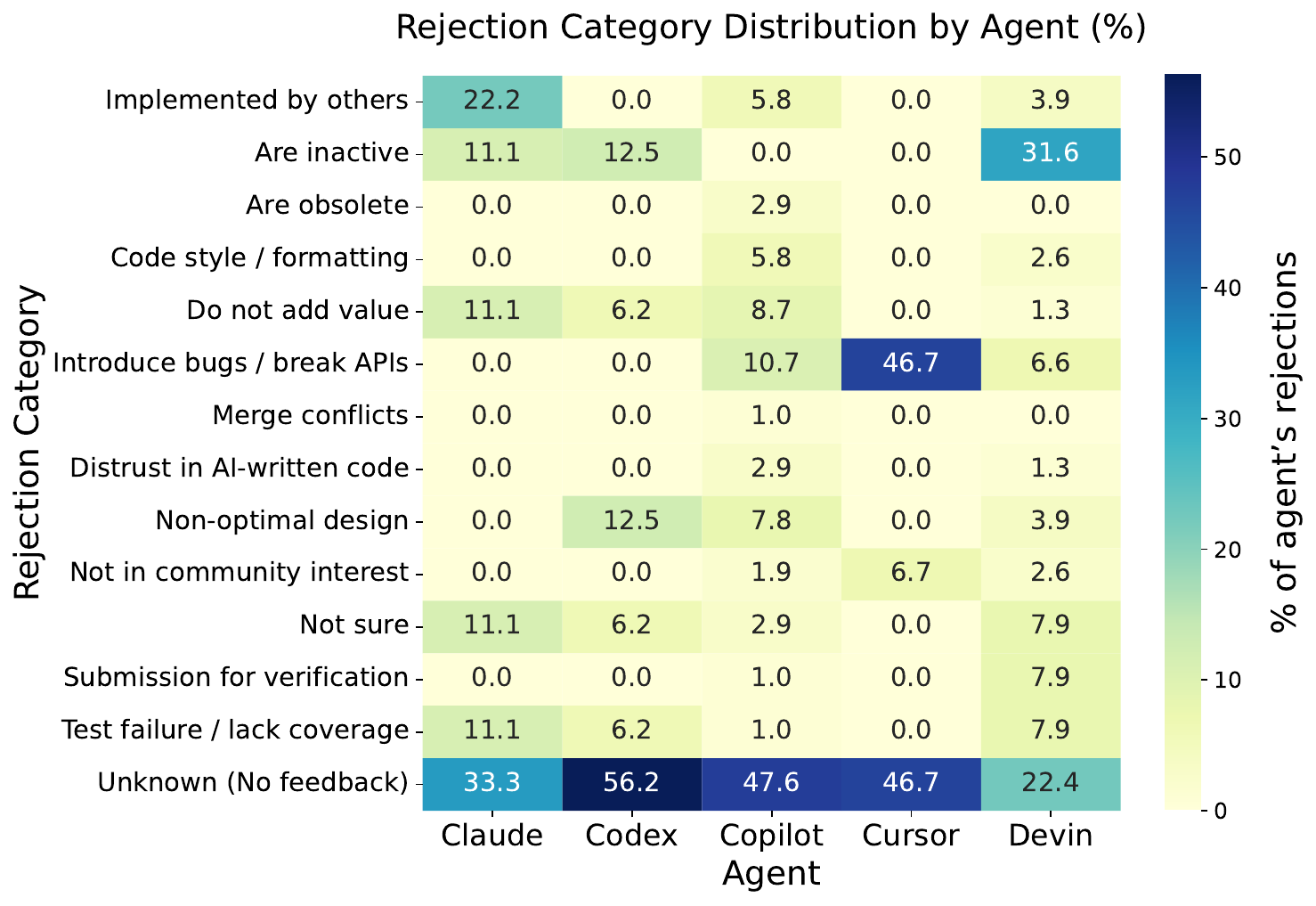}

\caption{Rejection reasons across AI agents}
\label{fig:rejection_heatmap}

\end{figure}

\begin{tcolorbox}[colback=gray!10,colframe=black,boxsep=1pt,left=2pt,right=2pt,top=2pt,bottom=2pt,boxrule=0.2pt] 
\textbf{Ans. to RQ4:} Most security-related AI PRs are rejected without feedback or due to inactivity, while the remaining rejections mainly cite technical issues such as bugs, design flaws, or missing tests. The distribution of rejection reasons varies across AI agents. 

\end{tcolorbox}

\section{Implications}
\label{sec:implications}

In this section, we discuss the implications of our findings organized for distinguished audiences such as researchers, practitioners, and tool builders working at the intersection of AI and software security.

\subsection{Implications for Researchers}

The findings suggest several directions for future research on agentic AI systems. Results from RQ1 show that a small group of vulnerabilities—such as regex inefficiencies, injection flaws, and path traversal—appear most frequently in security patches produced by agentic AI. 
This pattern suggests that these systems continue to face challenges in secure input handling and string processing. 
Prior work has shown that AI generated code often contains a higher number of high risk vulnerabilities than human written code~\cite{cotroneo2025human}, but most of that research focuses on general code generation rather than security fixes. 
Future studies should directly compare AI and human authored security patches to determine whether these recurring weaknesses are unique to agentic systems.

RQ3 and RQ4 reveal a gap between benchmark evaluation and practical review. 
Message quality does not appear to affect acceptance or review speed, while rejection analysis uncovers new reasons—such as \textit{code style or formatting} and \textit{test failure or insufficient coverage}—that have not been observed in studies of general pull requests. 
Existing benchmarks emphasize functional correctness~\cite{chon2024functional, paul2024benchmarks, liu2023your}, yet maintainers often evaluate AI contributions based on testing completeness and code quality. 
Future research may focus on designing evaluation frameworks that reflect these broader review expectations and investigate how reviewers perceive and trust agentic contributions. 
Recent evidence shows that repeated AI code generation can degrade rather than improve software security~\cite{shukla2025security}. 
Together, these insights underscore the need to examine the long-term reliability and security behavior of agentic systems as they operate across multiple commits and review cycles.

\subsection{Implications for Developers}

The results expose an imbalance in how reviewers handle agentic AI contributions.
RQ1 shows that some AI PRs introduce serious security flaws, such as command injection, yet they still pass review and are merged. In contrast, RQ4 reveals that other contributions are rejected for minor quality issues such as inconsistent style or missing test coverage. This contrast suggests that current review workflows do not consistently align review effort with security risk. Developers and maintainers may need clearer triage mechanisms that separate high-risk submissions from low-impact ones. Lightweight static checks or automated severity scoring could help identify critical flaws while allowing minor issues to be resolved through iteration.

The large share of rejections without feedback also points to a breakdown in the feedback cycle. In our dataset, 38.8\% of security PRs were closed without explanation, compared to 63.7\% in general agentic PRs reported by Watanabe et al.~\cite{watanabe2025use}. Although security-focused PRs receive somewhat more feedback, the absence of reasons still limits transparency and prevents both developers and AI builders from learning which kinds of errors trigger rejection. Even a brief structured tag, such as “risk,” “test,” or “design,” could help reviewers communicate intent and support continuous improvement of AI development tools.

\subsection{Implications for Coding Agent Builders}

The rejection patterns reveal practical directions for improving coding agents.
RQ4 shows that each system faces distinct challenges in producing acceptable security patches. Copilot often generates pull requests that fail due to design flaws or bugs. Devin tends to submit changes to inactive repositories that are later closed by automated bots. Cursor frequently introduces regressions, while Codex submissions are more often closed without feedback. These results suggest that future agent design should include better validation of target repositories, stronger self-review mechanisms, and clearer output provenance. Improving these capabilities would help align agent behavior with the expectations of human reviewers.

Beyond system tuning, our findings point to design principles for coding agents. Many rejections stem not only from technical flaws but also from limited transparency in the contribution. Reviewers often face uncertainty about the intent behind a design choice or the adequacy of testing. Coding agents could mitigate this uncertainty by generating structured review artifacts, such as short explanations of design rationale, evidence of test execution, and confidence scores on patch safety. Providing this context would reduce reviewer effort and improve trust in automated code contributions.

\section{Threats to Validity}
\label{sec:threats}

\textbf{Internal Validity:}
A key concern is the identification of security related PRs. Although we combine keyword lists from three prior studies to make the filter as comprehensive as possible and to minimize false positives, no list is exhaustive. We therefore use keywords only for initial filtering and rely on the \texttt{gemini-2.0-flash} API to refine the final set of security-related PRs. Because the API accuracy is uncertain, we manually review a stratified sample and obtain a Cohen’s $\kappa$ of 0.62, which shows moderate agreement between annotators. The manual validation is performed only on the 675 PRs labeled as security-related and not on the original 1,047 candidates; therefore, it captures false positives but not false negatives.
Another threat is tool accuracy. Semgrep may report false positives or miss vulnerabilities. Using Semgrep with default rules and relying on a single tool may miss certain vulnerabilities and limit CWE coverage, which can affect the completeness of our findings. Finally, two annotators assign rejection categories. Although they resolve disagreements through discussion, some subjectivity remains.

\noindent \textbf{Construct Validity:}
Our measures represent proxies for broader concepts. We treat Semgrep alerts as indicators of potential security issues, although not all alerts correspond to exploitable vulnerabilities. Rejection reasons depend on maintainer comments, which are sometimes missing or vague, leading to a large “Unknown” category. Commit message quality is estimated with the C-Good model, which captures the presence of “What” and “Why” but may not fully reflect reviewer perceptions of message usefulness.

\noindent\textbf{External Validity:}
Our dataset includes 675 AI-generated security PRs from GitHub repositories with more than 100 stars. This focus favors active open-source projects and excludes smaller or private repositories. The analysis is limited to five AI agents (Codex, Copilot, Devin, Cursor, and Claude) and PRs created before August 2025. Findings may not generalize to other agents, ecosystems, or future practices. Nonetheless, the selected projects span multiple domains and languages, reflecting trends in widely used repositories.

\noindent\textbf{Conclusion Validity:}
We use regression models to study factors related to PR latency and acceptance. Preprocessing steps such as log transformation, KNN imputation, and AutoSpearman reduce noise and multicollinearity, but model assumptions may not always hold. The sample size provides sufficient statistical power for overall analysis, though smaller subsets have reduced reliability. Our results describe statistical associations rather than causal relationships, and should be interpreted accordingly.

\section{Conclusion}
\label{sec:conclusion}

This paper examines how AI coding agents contribute security pull requests in open source projects. We analyze their vulnerability patterns, review outcomes, and rejection reasons to understand how these systems interact with human review processes. The findings show that AI authored security patches often repeat a narrow set of weaknesses such as regular expression inefficiencies, injection flaws, and path traversal. Despite such flaws, many patches are still merged, which points to gaps in review practices. Rejections, on the other hand, are frequently linked to social or process related factors such as inactivity or missing test coverage rather than to clear technical faults. Moreover, commit message quality, which strongly influences human authored contributions, does not appear to affect AI submissions.

These results show that current review processes are not well aligned with the behavior of AI coding agents. Reviewers often miss recurring security weaknesses in merged patches while rejecting other changes for minor issues. This mismatch calls for review workflows that better separate high-risk flaws from low-impact quality issues and for automated aids that highlight security-critical edits within AI pull requests. Future work can compare AI and human security fixes directly, study how reviewers form trust in AI code, and build benchmarks that reflect both technical and social aspects of review. Strengthening these areas will help make AI contributions more reliable and secure in open source software.

\section*{Data Availability}

All datasets and analysis scripts used in this study are publicly available as a replication package~\cite{dataset}.

\bibliographystyle{unsrt}  

\bibliography{references}

\end{document}